\documentclass[11pt]{article}
\usepackage{moriond,epsfig}

\def\Journal#1#2#3#4{{#1} {\bf #2}, #3 (#4)}

\def\be{\begin{equation}}
\def\ee{\end{equation}}
\def\bea{\begin{eqnarray}}
\def\eea{\end{eqnarray}}

\begin{document}
\vspace*{4cm}
\title{ALICE COMMISSIONING: GETTING READY FOR PHYSICS}

\author{C. LIPPMANN}

\address{PH Division, CERN, 1211 Geneva 23, Switzerland}

\maketitle\abstracts{
  ALICE is a general-purpose heavy-ion detector at the CERN LHC. We
  describe the commissioning activities in the years 2007-2009 and the
  status of the different subsystems at the time of this conference.
}

\vspace*{-0.5cm}
\section{Introduction}
\vspace*{-0.2cm}

ALICE (A Large Ion Collider Experiment) is the dedicated heavy-ion experiment
at the Large Hadron Collider (LHC) in Geneva\cite{proposal}. It is designed
to study the physics of strongly interacting matter at extreme energy
densities by analysing the collisions of lead nuclei at $\sqrt{s}=5.5$\,TeV
per nucleon pair, using as probes the hadrons, electrons, muons and
photons that are produced in the collisions\cite{ppr}. ALICE must
offer excellent particle identification (PID) in a large momentum range.
As compared to proton collisions at the LHC, the charged track multiplicities
will be extraordinary for the lead collisions\footnote{ALICE was optimised
  for a charged-track multiplicity $dN_{ch}/d\eta\leq 8000$.}, but the
momenta of the produced particles will be rather low. As a consequence,
precision tracking capabilities over a large momentum range
(100\,MeV/c\,$<p<100$\,GeV/c), a low material budget and a rather low
magnetic field\footnote{The magnetic field of the ALICE solenoid magnet
  (L3) will be in the range $0.2$\,T$\leq B\leq 0.5$\,T.}
are required. A very high granularity detector was chosen, which is
however---at least when compared to the other large LHC experiments---limited
in readout speed.

\vspace*{-0.4cm}
\section{ALICE Commissioning in 2007-2009}\label{com0709}
\vspace*{-0.2cm}

The installation and commisioning of ALICE subdetectors continued well into
the year 2008. Two global runs were organised in late 2007 and in early 2008
in order to integrate the individual subdetectors in the global experiment
control system. From May to October 2008 the third global run
took place, including 24 hour operation over many months, and calibration data
taking for the various subdetectors. In this period, the first particles from
the LHC reached the experiment on 15th June. During the LHC injection tests on
8th and 24th August the first beam-related events were seen in ALICE. The third
global run then started on 5th May 2008. It was intended to converge into data
taking with beam collisions, and on 10th September the first beams were actually
circulated in the LHC. Unfortunately, on 19th September a serious fault damaged
a number of LHC magnets and the beams stopped. The data taking however continued
until 20th October with calibration triggers. The LHC will not see beam
again before September 2009. The resulting extended shutdown period is used
for installation of more detector modules (TRD, PHOS and EMCAL, see below), and
for upgrades and maintenance.

\vspace*{-0.4cm}
\section{ALICE Experimental Setup and Status of the Subdetectors}
\vspace*{-0.2cm}

ALICE consists of a central barrel part for the measurement of hadrons,
leptons and photons, and a forward muon spectrometer. The central part
is embedded in a large solenoid magnet that is reused from the L3
experiment at LEP. More detailed descriptions of the detector systems
can be found elsewhere\cite{alice}. Their status at the time of this
conference (March 2009) is described in the following.

\vspace*{-0.2cm}
\subsection{ITS} \label{setupstatus:its}
\vspace*{-0.2cm}

\begin{figure}
  \centering
  \includegraphics[width=16cm]{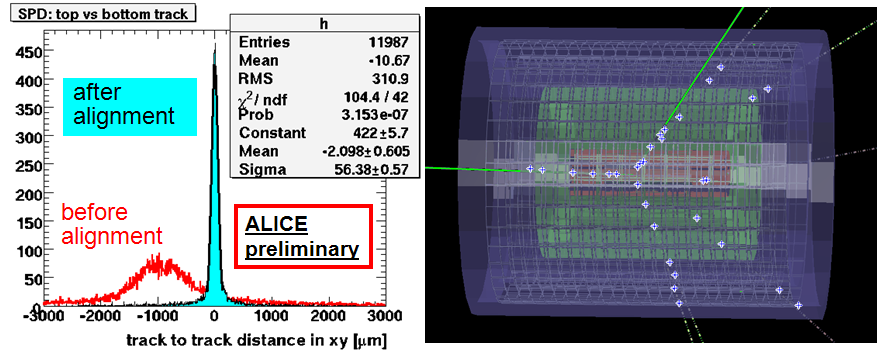}
  \vspace{-0.8cm}
  \caption{Left: SPD resolution after alignment with cosmic tracks. We
    compare the upper and lower halfs of cosmic tracks reconstructed in the SPD.
    Right: Tracks from a common vertex reconstructed in the ITS. They indicate
    a beam induced collision. The data was taken during LHC beam circulation
    on 11th September 2008.}
  \label{fig:its}
\end{figure}

The {\it Inner Tracking System} consists of three different silicon
detector technologies with two layers each: High resolution silicon
pixel (SPD), drift (SDD) and strip (SSD) detectors. The system is
fully installed and commissioned. In general, the whole ITS was
recording data during the LHC injection tests in august 2008 and
during the first beam circulations. Preliminary results of the
internal SPD alignment based on a sample of cosmic tracks and a
beam induced event display are shown in Fig. \ref{fig:its}. The
achived performance compares well to simulations; with the
residual misalignment the design value for the primary vertex
resolution (100\,$\mu$m for proton colisions) is expected to be
reached.
 
The SPD can provide a trigger decision at the first level. Each of the
1200 pixel chips transmits a digital pulse, when one or more of the
pixels in the matrix are hit. Different simple and fast algorithms
can be used to provide triggers based on multiplicities. ALICE is the
only LHC experiment including the vertex detector in the first trigger
decision from startup.

\vspace*{-0.2cm}
\subsection{TPC} \label{setupstatus:tpc}
\vspace*{-0.2cm}

\begin{figure}
  \centering
  \includegraphics[width=16cm]{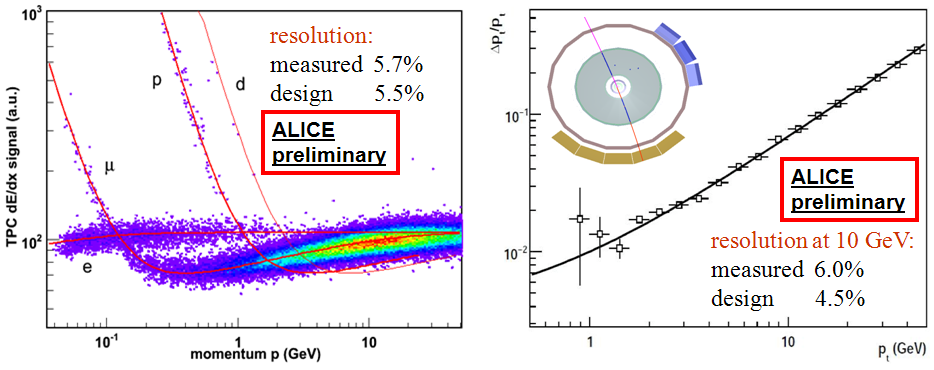}
  \vspace{-0.8cm}
  \caption{Performance of the TPC measured with cosmic trigger as a
    function of momenta. Left: d$E$/d$x$ spectra. Right: Transverse
    momentum resolution. The inlay shows a typical track used to derive
    the momentum resolution (by comparing the upper and lower half).}
  \label{fig:tpc}
\end{figure}

The main tracking detector in the central barrel is the
{\it Time Projection Chamber}, offering also momentum measurement and
PID. Its main advantage is the unequalled granularity (560\,Mio pixels).
However, some of the requirements are very challenging: The TPC has to
cope with very high charged track multiplicities, but it also has to
be read out very fast: 1\,kHz (a few 100\,Hz) for proton (lead)
collisions. The TPC was running continuously for many months in 2008
and recorded $60\times 10^6$ calibration events. The first round of
calibrations is completed and reveals a performance already approaching
the design values (Fig. \ref{fig:tpc}). Very good temperature
homogenisation in the drift volume\footnote{This is necessary in order
  to limit changes in the drift velocity.} was achieved, close to the
design value ($\Delta T < 0.1$\,K).

\vspace*{-0.2cm}
\subsection{ACORDE} \label{setupstatus:acorde}
\vspace*{-0.2cm}

The {\it Alice COsmic Ray DEtector} is an array of plastic scintillator
paddles mounted on the L3 magnet. It serves as a cosmic ray trigger
at the first trigger level. Only cosmic muons with energies $\geq 10$\,GeV reach the
ACORDE, at rates $\leq 5$\,Hz/m$^2$. The system is fully commissioned
and is used heavily in the commissioning of other ALICE detectors.

\vspace*{-0.2cm}
\subsection{Outer Central Detectors} \label{setupstatus:outer}
\vspace*{-0.2cm}

\begin{figure}
  \centering
  \includegraphics[width=16cm]{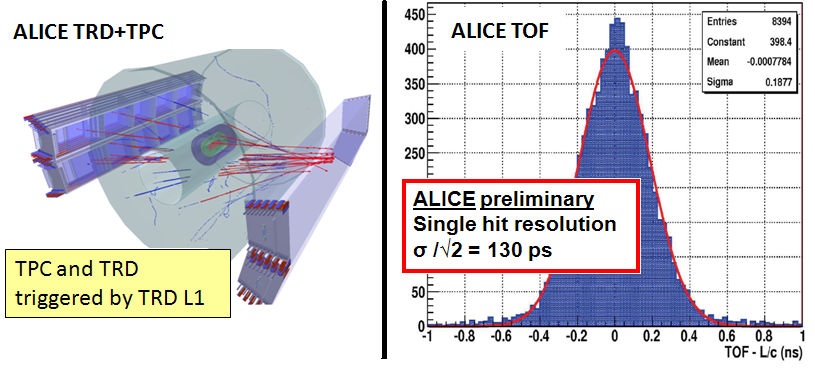}
  \vspace{-0.8cm}
  \caption{Left: An event display in the TPC and TRD. The readout was
    triggered by the TOF (first trigger level) and confirmed by the
    TRD trigger (second level). Right: Histogram of the measured
    flight time between two TOF modules. We subtract $L/c$, with $L$ the
    length of the trajectory, and $c$ the speed of light.}
  \label{fig:tof-trd}
\end{figure}

The {\bf TRD} (Transition Radiation Detector) is not yet fully installed
(4 out of 18 TRD modules during 2008). For the run in 2009/2010 up to 8
modules will be installed. The TRD trigger contribution (second level) on
high momentum particles was sucessfully used in 2008 (Fig. \ref{fig:tof-trd}).
The {\bf TOF} (Time Of Flight) implements Multigap Resistive Plate Chambers
to obtain a very good system time resolution sufficient to provide
PID in the intermediate momentum range. All 18 TOF modules are
installed and fully commissioned. During 2008 a system time resolution of
130\,ps was achieved in the experiment (Fig. \ref{fig:tof-trd}).
This number has to be compared to a design value of around 100\,ps and is
expected to improve significantly following more detailed calibrations.
The {\bf HMPID} (High Momentum Particle IDentification) extends the
useful PID range ($\pi$/$K$, $K$/$p$). It was the first detector to be
installed in the L3 magnet and is fully commissioned.
The {\bf PHOS} (PHOton Spectrometer) provides PID ($\gamma$, $\pi^0$ and
$\eta$) and contributes to the first level trigger decision. One out of 5
modules was installed in 2008. For the run in 2009/2010 up to 3 modules
will be installed.
The aim of the {\bf EMCAL} (ElectroMagnetic CALorimeter) is to do high
momentum jet physics. The project was approved only in the end of 2007 and
up to 4 out of 12 modules will be installed for the run in 2009/2010.

\vspace*{-0.2cm}
\subsection{Muon Spectrometer} \label{setupstatus:muon}
\vspace*{-0.2cm}

The muon spectrometer will study the complete spectrum of heavy
quarkonia (J/$\Psi$, $\Psi'$ and the $\Upsilon$ family) via their decay
in the $\mu^+\mu^-$ channel. It consists of 5 tracking stations (2 planes
of cathode pad chambers each), 2 trigger stations (2 planes of Resistive
Plate Chambers each), a dipole magnet, and two absorbers. It covers the
pseudorapidity interval $2.5\leq\eta\leq 4$ and the whole system is now
fully commissioned.

\vspace*{-0.2cm}
\subsection{Forward Detectors} \label{setupstatus:forward}
\vspace*{-0.2cm}

A number of small and specialized detector systems in the forward
region are used for triggering or to measure global event
characteristics. All of them were already fully installed and
commissioned in 2008.

The {\bf FMD} (Forward Multiplicity Detector) consists of 3 planes of
silicon strip detectors to measure charged particle multiplicities.
During the LHC injection tests and beam circulation the FMD was
in general recording data and operated as expected at hit
densities of 10 to several $10^3$ charged particles per cm$^2$.
The {\bf T0} detector serves as time reference for TOF ($\sim 30$\,ps
time resolution) and provides vertex measurement.
The {\bf V0} detector provides a centrality trigger at the first level,
serves as a luminosity monitor and can be used to reject beam-gas
events. During the LHC injection tests and beam circulation the
V0 was in general recording data and for the beam-induced events
clear correlations between the measured signal charge and, e.g. the
number of fired chips in the SPD were observed.
The {\bf ZDC} (Zero Degree Calorimeter) is composed of four
calorimeters, two each to detect protons and neutrons. They are
located 115 meters away from the interaction point on both sides,
exactly along the beam line.
The {\bf PMD} (Photon Multiplicity Detector) measures multiplicity
and spatial distributions of photons in the forward pseudorapidity
region.

\vspace*{-0.2cm}
\section{Summary \& Conclusions}
\vspace*{-0.2cm}

Where possible the detector performance was evaluated using cosmic muons
and the first few beam particles delivered by the LHC. All installed
detectors are fully commissioned and shown to be performing close to their
specifications. The current shutdown period is used to install additional
detectors.

The ALICE experiment is ready for recording the first proton-proton
collisions in the LHC, expected in late 2009, and the collaboration
eagerly awaits the first heavy-ion collisions at the end of the upcoming
run.

\end{document}